\documentclass{ws-mpla}

\begin{document}

\markboth{A.V.Gladyshev, D.I.Kazakov, M.G.Paucar} {Light Stops in
the MSSM Parameter Space}

\catchline{}{}{}{}{}

\title{LIGHT STOPS IN THE MSSM PARAMETER SPACE}

\author{\footnotesize A.V.GLADYSHEV$^{1,2}$, D.I.KAZAKOV$^{1,2}$, M.G.PAUCAR$^1$}

\address{$^1$ Bogoliubov Laboratory of
Theoretical Physics, Joint Institute for Nuclear Research,\\
141980, 6 Joliot-Curie, Dubna, Moscow Region, Russian
Federation\\
gladysh@theor.jinr.ru, kazakovd@theor.jinr.ru, manuel@theor.jinr.ru}


\address{$^{2}$Institute of Theoretical and Experimental Physics,\\
117218, 25 Bolshaya Cheremushkinskaya, Moscow, Russian Federation}

\maketitle

\pub{Received (Day Month Year)}{Revised (Day Month Year)}

\begin{abstract}
We consider the regions of the MSSM parameter space where the top
squarks become light and even may be the LSP. This happens when the
triple scalar coupling $A$ becomes very big compared to $m_0$. We
show that in this case the requirement that the LSP is neutral
imposes noticeable constraint on the parameter space excluding low
$m_0$ and $m_{1/2}$ similar to constraint from the Higgs mass limit.
In some cases these constraints overlap. This picture takes place in
a wide region of $\tan\beta$. In a narrow band close to the border
line the stops are long-lived particles and decay into quarks and
neutralino (chargino). The cross-section of their production at LHC
via gluon fusion mechanism in this region may reach a few pb.

\end{abstract}

\ccode{PACS Nos.: 12.60.Jv, 14.80.Ly}

\section{Introduction}

Preparing for SUSY discovery at LHC one faces the problem of well
defined predictions since the variety of models and scenarios open
up a wide range of possibilities~\cite{MSSM1,MSSM2,MSSM3,MSSM4,MSSM5,MSSM6}.
Due to the absence of a SUSY
golden mode, one has to explore the parameter space looking for high
cross-sections, low background processes, typical missing energy
events, etc. All these possibilities are realized within some
mechanism of SUSY breaking (mSUGRA and gauge mediation are the most
popular) and depend on particular choice of a region in SUSY
parameter space~\cite{CMSSM1,CMSSM2,CMSSM3,CMSSM4,CMSSM5,CMSSM6,CMSSM7,CMSSM8}.

Besides commonly accepted benchmark points which are widely
discussed in the literature~\cite{benchmark1,benchmark2,benchmark3,benchmark4}
there still exist some exotic regions in
parameter space where unusual relations hold and one can expect
interesting phenomena. In particular, in Ref.~\cite{us} we
considered the so-called coannihilation region in mSUGRA parameter
space where one can have long-lived staus which can decay at some
distance from the collision point or even fly through detector. This
region of $m_0 - m_{1/2}$ plane exists for all values of $\tan\beta$
and moves towards higher values of $m_0$ and  $m_{1/2}$  with
increase of the latter. However, the area where one can have
long-lived particles is very narrow (for each $\tan\beta$) and needs
severe fine-tuning.

Here we explore another region of parameter space which appears only
for large negative scalar triple coupling $A_0$ and is distinguished by the
light stops. On the border of this region, in full analogy with the
stau coannihilation region, the top squark becomes the LSP and near
this border one might get the long-lived stops. Below we discuss
this possibility in detail and consider also its phenomenological
consequences for the LHC.

\section{Constraints on the MSSM Parameter Space for Large Negative Values of~$A$}

In what follows we consider the MSSM with gravity mediated
supersymmetry breaking and the universal soft terms. We thus have
the parameter space defined by $m_0, m_{1/2},$ $A, \tan\beta$ and we
take the sign of $\mu$ to be positive motivated by contribution to
the anomalous magnetic moment of muon~\cite{Anom1,Anom2,Anom3,Anom4,Anom5,Anom6,Anom7,Anom8}. Imposing the constraints
like: i) the gauge couplings unification~\cite{Unif1,Unif2,Unif3}, ii) neutrality of the LSP~\cite{LSP1,LSP2},
iii) the Higgs boson and SUSY mass experimental limits~\cite{higgslimits1,higgslimits2,higgslimits3}, iv)
radiative electroweak symmetry breaking, we get the allowed region
of parameter space. Projected to the $m_0 - m_{1/2}$ plane this region
depends on the values of $\tan\beta$ and $A$. In case when $A$ is
large enough the squarks of the third generation, and first of all
stop, become relatively light. This happens via the see-saw
mechanism while diagonalizing the stop mass matrix
\begin{equation} \label{stopmat}
\left(\begin{array}{cc} \tilde m_{tL}^2& m_t(A_t-\mu\cot \beta )
\\ m_t(A_t-\mu\cot \beta ) & \tilde m_{tR}^2 \end{array}  \right),
\nonumber
\end{equation}
where
\begin{eqnarray*}
  \tilde m_{tL}^2&=&\tilde{m}_Q^2+m_t^2+\frac{1}{6}(4M_W^2-M_Z^2)\cos
  2\beta ,\\
  \tilde m_{tR}^2&=&\tilde{m}_U^2+m_t^2-\frac{2}{3}(M_W^2-M_Z^2)\cos
  2\beta .
\end{eqnarray*}
The off-diagonal terms increase with $A$, become large for large
$m_q$ (that is why the third generation) and give negative
contribution to the lightest squark mass defined by minus sign in
eq.(\ref{stop}).
\begin{equation}
\tilde{m}^2_{1,2}= \frac 12 \left(\tilde m_{tL}^2+\tilde m_{tR}^2
\pm \sqrt{ (\tilde m_{tL}^2-\tilde m_{tR}^2)^2+4m_t^2(A_t-\mu
\cot\beta})^2\right).
 \label{stop}
\end{equation}

\begin{figure}[tp]
\centerline{\psfig{file=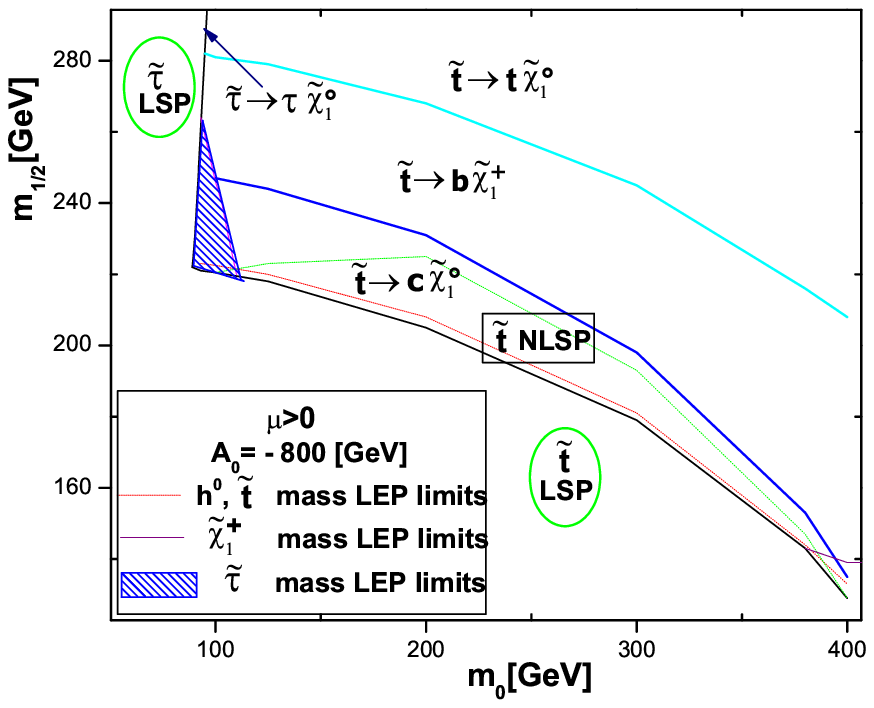,width=0.7\textwidth}} \vspace*{8pt}
\centerline{\psfig{file=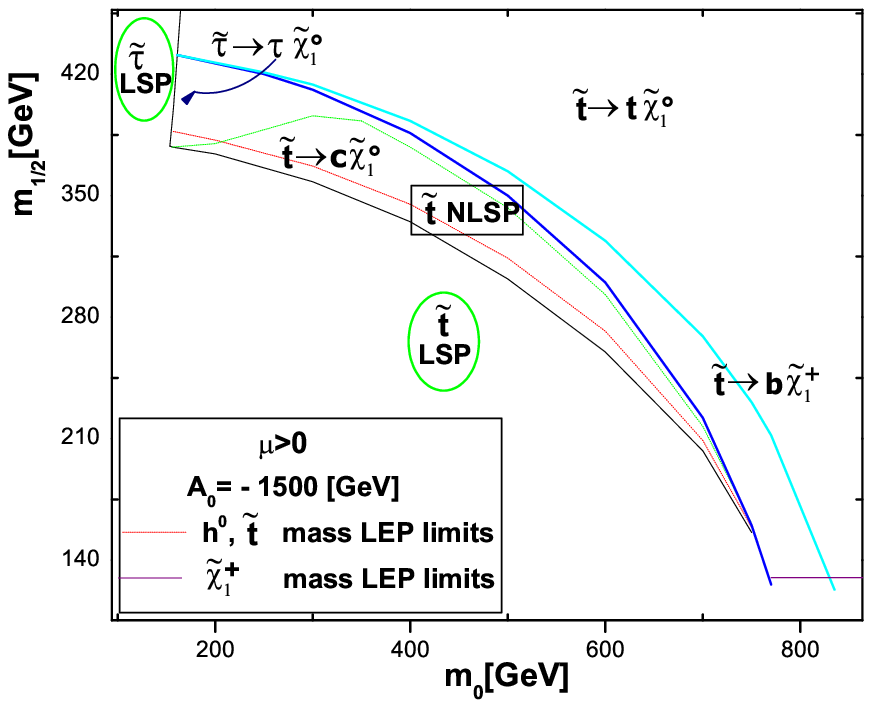,width=0.7\textwidth}} \vspace*{8pt}
\caption{Allowed region of the mSUGRA parameter space for
$
A_0= -800, -1500
$ GeV  and $\tan\beta=10$.  At the left from the
border stau is an LSP, below the border stop is the LSP. The dotted
line is the LEP Higgs mass limit. Also shown are the contours where
various stop decay modes emerge.
\protect\label{fig:fig1}}
\end{figure}

\begin{figure}[tp]
\centerline{\psfig{file=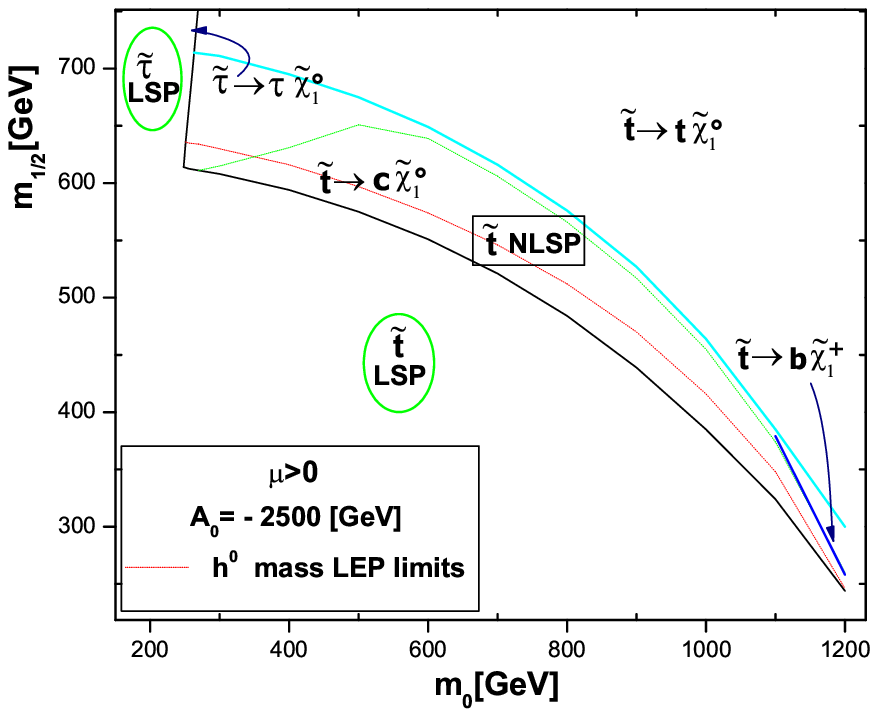,width=0.7\textwidth}} \vspace*{8pt}
\centerline{\psfig{file=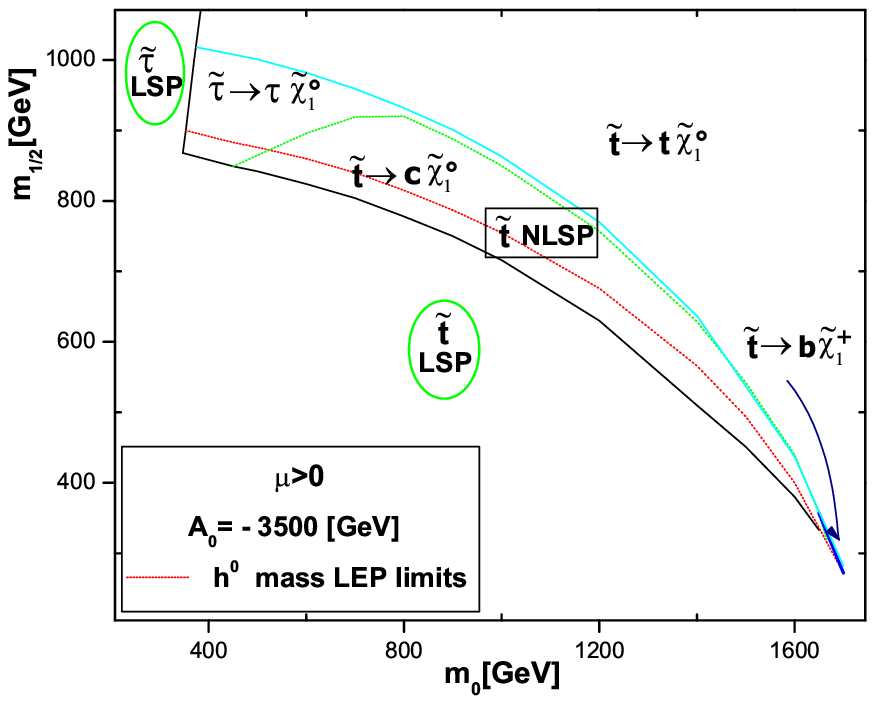,width=0.7\textwidth}} \vspace*{8pt}
\caption{Allowed region of the mSUGRA parameter space for
$
-2500, -3500
$ GeV and $\tan\beta=10$.  At the left from the
border stau is an LSP, below the border stop is the LSP. The dotted
line is the LEP Higgs mass limit. Also shown are the contours where
various stop decay modes emerge.
\protect\label{fig:fig2}}
\end{figure}

Hence, increasing $|A|$ one can make the lightest stop as light as
one likes it to be and even make it the LSP. The situation is
similar to that with stau for small $m_0$ and large $m_{1/2}$ when
stau  becomes the LSP. For squarks it takes place for low $m_{1/2}$
and low $m_0$.  One actually gets the border line where stop becomes
the LSP. The region below this line is forbidden.  It  exists only
for large negative $A$, for small $A$ it is completely ruled out by
the LEP Higgs limit.

\begin{figure}[tb]
\centerline{\psfig{file=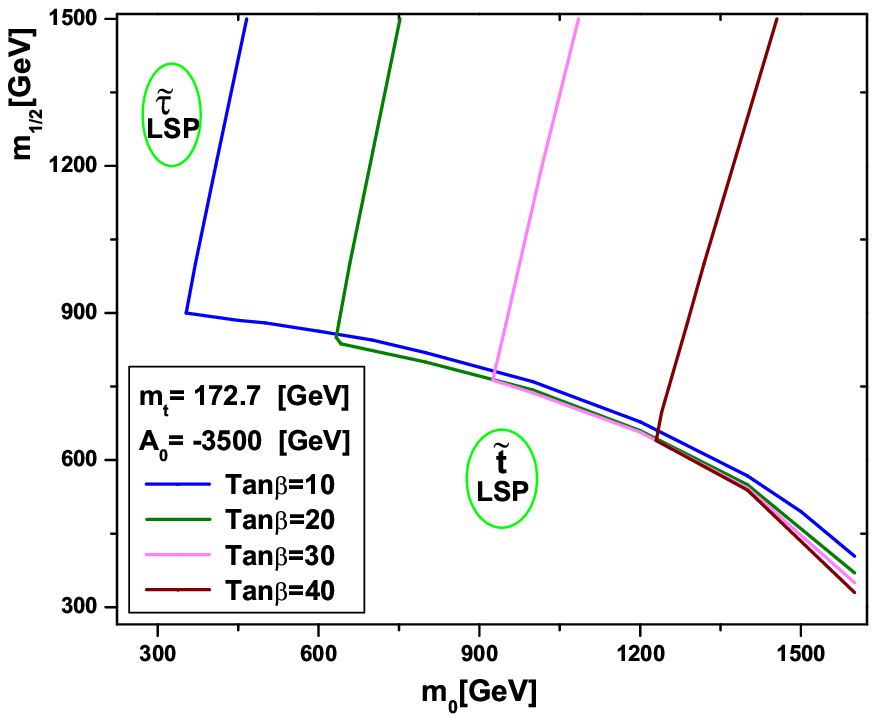,width=0.8\textwidth}} \vspace*{8pt}
\caption{Stau and stop constraints in the $m_0 - m_{1/2}$ plane for $A_0=-3500$ GeV
and different values of $\tan\beta$.
\protect\label{fig:fig3}}
\end{figure}

It should be noted that in this region  one  gets not only the light
stop, but also the light Higgs, since the radiative correction to the Higgs
mass is proportional to the log of the stop mass.
The stop mass boundary is close to
the Higgs mass one and  they may overlap for intermediate values of
$\tan\beta$. We show the projection of SUSY parameter space to the
$m_0 - m_{1/2}$ plane in Figs.~\ref{fig:fig1} and~\ref{fig:fig2} for different values of $A$
and fixed $\tan\beta$. To calculate it we use the ISAJET v.7.64
code~\cite{ISAJET}.

One can see that when $|A|$ decreases the border line moves down and
finally disappears. On the contrary, increasing $|A|$ one gets
larger forbidden area and the value of the stop mass at the border
increases.

Changing $\tan\beta$ one does not influence the stop border line,
the only effect is the shift of stau border line. It moves  to the
right with increase of $\tan\beta$ as shown in Fig.\ref{fig:fig3}, so
that the whole forbidden area increases and covers the left bottom
corner of the $m_0 - m_{1/2}$ plane.

It should be mentioned that the region near the border line is very
sensitive to the SM parameters, a minor shift in $\alpha_s$ or $m_t$
and $m_b$ leads to noticeable change of spectrum as can be seen from
comparison of different codes at~\cite{kraml1,kraml2,kraml3}.

The other constraint that is of interest in this region is the relic
density one. Given the amount of the Dark matter from WMAP
experiment~\cite{WMAP1,WMAP2} one is left with a narrow band of allowed
region which goes along the stau border line, then along the Higgs
limit line and then along the radiative symmetry breaking line. In
the case of light stop when it is almost degenerate with the
lightest chargino and neutralino, when calculating the relic density
one has to take into account not only the annihilation of two stops,
but also the coannihilation diagrams. There are two side processes:
stop chargino annihilation and stop neutralino annihilation.
Calculating the relic density with the help of MicrOmegas
package~\cite{micromegas1,micromegas2} one finds that again it is very sensitive
to the input parameters, however, since the stop border line is very
close to the Higgs one, the relic density constraint may be met here
fitting $A$ and/or $\tan\beta$.

\section{Phenomenological Consequences of the Light Stop Scenario}

\begin{figure}[htb]
\centerline{\psfig{file=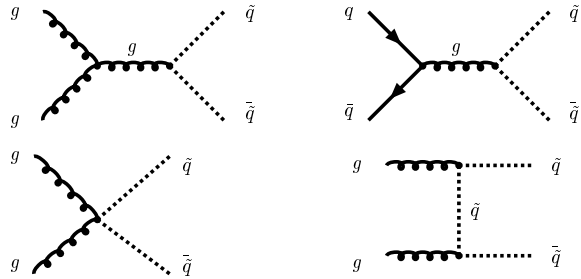,width=0.6\textwidth}} \vspace*{8pt}
\caption{Main stop production diagrams at LHC.
\protect\label{fig:fig4}}
\end{figure}

The phenomenology of the discussed scenario is of great interest
at the moment since the first physics results of the coming LHC are
expected in the nearest future. Light stops could be produced already
during first months of its operation~\cite{stopprod1,stopprod2}. The main diagrams of pair stop
production (as well as other type of squarks) are presented in Fig.~\ref{fig:fig4}.
A single stop production via weak interactions is also possible.

\begin{figure}[tp]
\centerline{\psfig{file=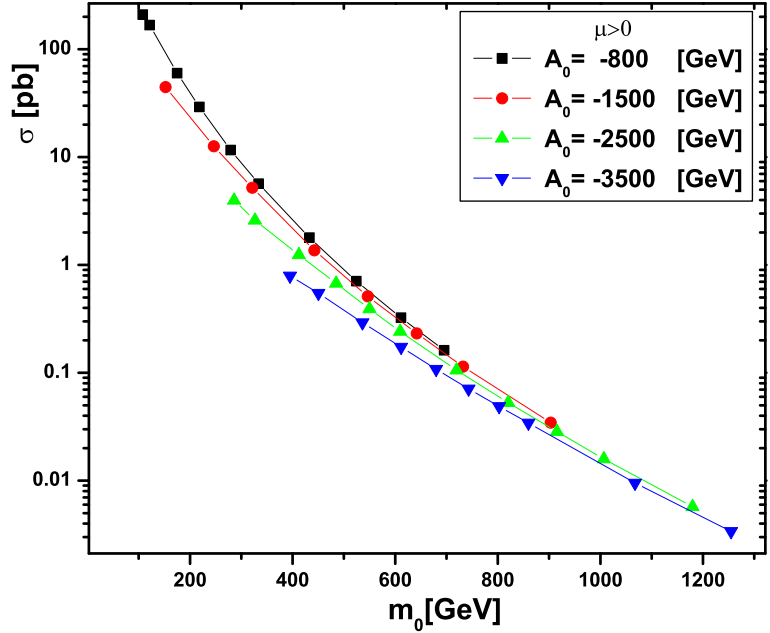,width=0.8\textwidth}} \vspace*{8pt}
\caption{Cross sections of the pair stop production as a function of the stop mass.
Different curves correspond to different values of $A_0$ parameter
($A_0=-800, -1500, -2500, -3500$ GeV).
\protect\label{fig:fig5}}
\end{figure}

Since stops are relatively light in our scenario, the production cross sections
are quite large and may achieve tens or even hundreds of pb for $m_{\tilde t}< 150$~GeV.
The cross sections and their dependence on the stop mass for different values of $|A|$
are shown in Fig.~\ref{fig:fig5}. As one expects they quickly fall down when the mass of
stop is increased. The range of each curve corresponds to the region in the $(m_0 - m_{1/2})$
plane where the light stop is the next-to-lightest SUSY particle, and the Higgs
and chargino mass limits are satisfied as well.
One may notice, that even for very large values of $|A|$ when stops become heavier than several hundreds GeV,
the cross sections are of order of few per cent of pb, which is still enough for detection
with the high LHC luminosity.

Being created the stop decay. There are several different decay modes depending on the
stop mass.
If stop is heavy enough  it decays to the bottom quark and the lightest chargino
($\tilde t \to b \tilde\chi^\pm_1$). However, for large values of $|A_0|$,
namely $A_0<-1500$~GeV the region where this decay takes place is getting smaller
and even disappear due to mass inequality $m_{\tilde t} < m_b + m_{\tilde\chi_1^\pm}$
(see right bottom corner in Fig.~\ref{fig:fig2}). In this case the dominant decay mode is
the decay to the top quark and the lightest neutralino ($\tilde t \to t \tilde\chi^0_1$).
Light stop decays to the charm quark and the lightest neutralino
($\tilde t \to c \tilde\chi^0_1$)~\cite{stopdecay}. The latter decay, though it is loop-suppressed,
has the branching ratio 100 \%.

\begin{figure}[th]
\centerline{\psfig{file=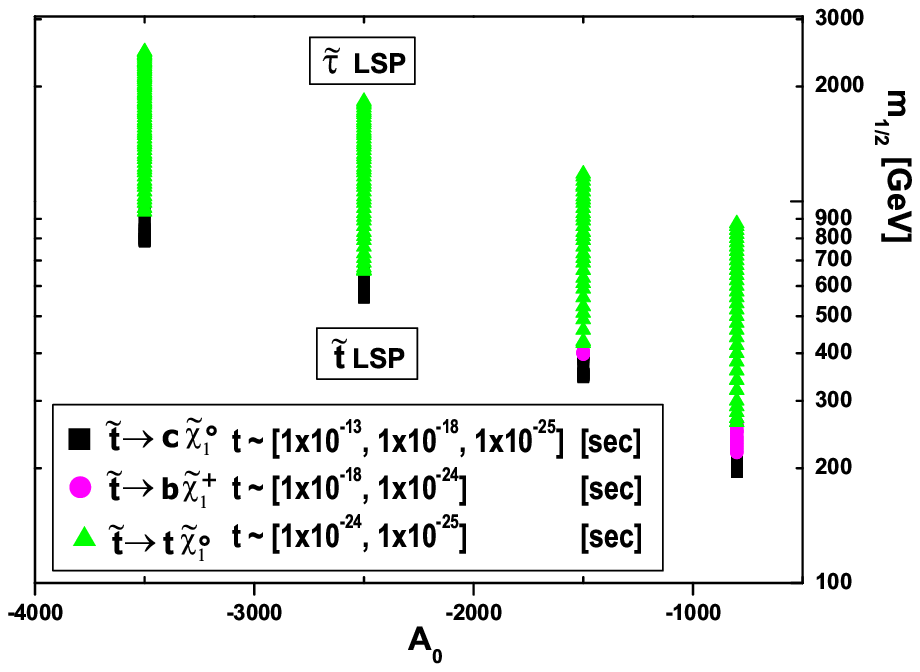,width=0.8\textwidth}} \vspace*{8pt}
\centerline{\psfig{file=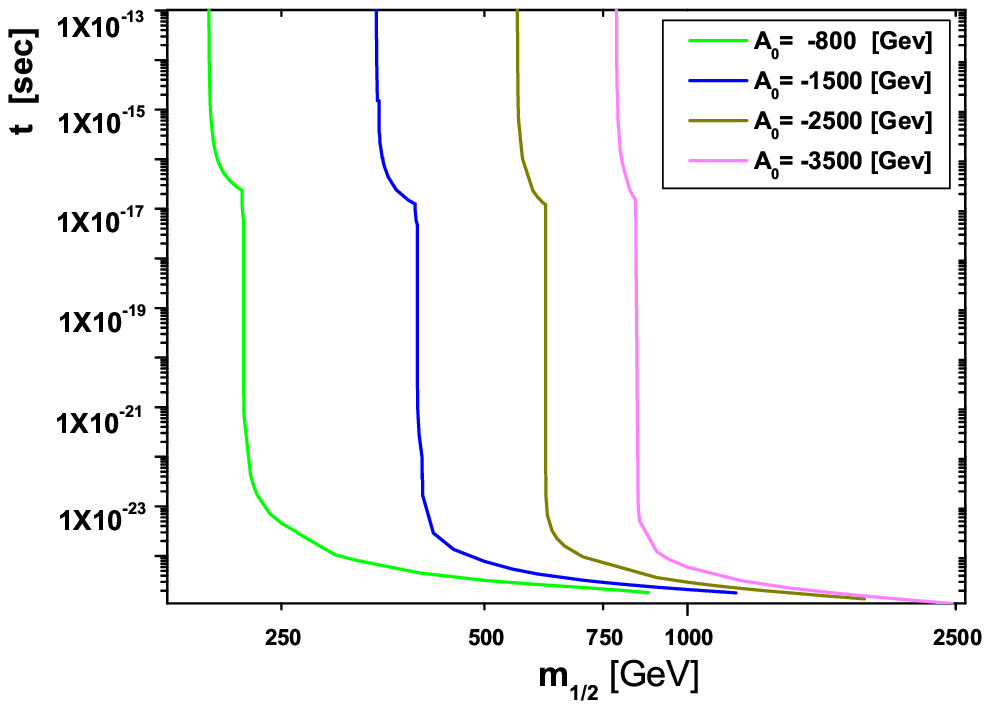,width=0.8\textwidth}} \vspace*{8pt}
\caption{Different decay modes of stops and corresponding lifetimes.
\protect\label{fig:fig6}
}
\end{figure}

In Fig.~\ref{fig:fig6} we show different allowed decay modes for different values of $|A_0|=800,1500,2500,3500$ GeV
as functions of the $m_{1/2}$ parameter. The values of $m_0$ were chosen in the middle of allowed regions
in Figs~\ref{fig:fig1} and~\ref{fig:fig2}, namely $m_0=250,450,650,1000$ GeV. One can see that for
small values of $m_{1/2}$ we are very close to the neutralino--stop border line and the only allowed
decay mode is $\tilde t \to c \tilde\chi^0_1$. With the increase of the $m_{1/2}$ the stop mass becomes larger
which opens the possibility of new decay modes ($\tilde t \to b \tilde\chi^\pm_1$,
and later $\tilde t \to t \tilde\chi^0_1$).

The bottom part of Fig.~\ref{fig:fig6} shows the stop
lifetimes for different values of $|A_0|$. The biggest lifetime corresponds to the
$\tilde t \to c \tilde\chi^0_1$ decay. Breaks on the curves correspond to switching on
the new decay mode. As one can see the lifetime cold be quite large in a wide area of the $A_0-m_{1/2}$
parameter space, even for heavy stops if $|A_0|$ is very big.

\section{Conclusion}

In this letter we have demonstrated that there is a possibility of
stop next-to-lightest supersymmetric particle. For large negative values of
the trilinear soft supersymmetry breaking parameter $A_0$ there exist a narrow band along the
line $m_{\tilde t}=m_{\tilde\chi^0_1}$ in the $m_0 - m_{1/2}$ plane where the
cross section of stop pair production is quite large at the LHC energy and stops have relatively large
lifetime. This may give interesting signatures, like secondary vertices inside the detector,
or even escaping the detector.
Another interesting possibility is a formation of so-called $R$-hadrons (bound states of supersymmetric particles).
This may happen if stops live longer than hadronisation time.

Experimental Higgs and chargino mass limits as well as WMAP relic density limit can be easily satisfied in our scenario.
However, the strong fine-tuning is required. Moreover, it is worth mentioning that light stops are favoured by
the baryon asymmetry of the Universe.

\section*{Acknowledgements}

Financial support from RFBR grant \# 05-02-17603 and grant of the Ministry of Education
and Science of the Russian Federation \# 5362.2006.2 is kindly acknowledged.


\begin{thebibliography}{40}
\bibitem{MSSM1}
    H.P. Nilles,
    \textit{Phys. Rept.} \textbf{110}, 1 (1984).
\bibitem{MSSM2}
    H.E. Haber and G.L. Kane,
    \textit{Phys. Rept.} \textbf{117}, 75 (1985).
\bibitem{MSSM3}
    H.E. Haber,
    \emph{Introductory Low-Energy Supersymmetry},
    Lectures given at TASI 1992, (SCIPP 92/33, 1993),
    \texttt{hep-ph/9306207}.
\bibitem{MSSM4}
    D.I. Kazakov,
    \emph{Beyond the Standard Model}, Lectures at the European School of
    High-Energy Physics 2004, \texttt{hep-ph/0411064}.
\bibitem{MSSM5}
    A.V. Gladyshev and D.I. Kazakov,
    \textit{Yadernaya Fizika} \textbf{70}, No.7, 1 (2007).
\bibitem{MSSM6}
    A.V. Gladyshev and D.I. Kazakov,
    \emph{Supersymmetry and LHC}, Lectures given at 9th International
    Moscow School of Physics and 34th ITEP Winter School of Physics,
    \texttt{hep-ph/0606288}
\bibitem{CMSSM1}
    L.E. Ib\'{a}\~{n}ez and G.G. Ross,
    \textit{Nucl. Phys. B} \textbf{368}, 3 (1992).
\bibitem{CMSSM2}
    G.G. Ross and R.G. Roberts,
    \textit{Nucl. Phys. B} \textbf{377}, 571 (1992).
\bibitem{CMSSM3}
    S. Kelley, J.L. Lopez and D.V. Nanopoulos,
    \textit{Phys. Lett. B} \textbf{274}, 387 (1992).
\bibitem{CMSSM4}
    V. Barger, M.S. Berger, P. Ohmann and R. Phillips,
    \textit{Phys. Lett. B} \textbf{314}, 351 (1993).
\bibitem{CMSSM5}
    V. Barger, M.S. Berger and P. Ohmann,
    \textit{Phys. Rev. D} \textbf{49}, 4908 (1994).
\bibitem{CMSSM6}
    P. Langacker and N. Polonsky,
    \textit{Phys. Rev. D} \textbf{49}, 1454 (1994).
\bibitem{CMSSM7}
    W. de Boer, R. Ehret and D.I. Kazakov,
    \textit{Z. Phys. C} \textbf{67} 647 (1995).
\bibitem{CMSSM8}
    W. de Boer \textit{et al.},
    \textit{Z. Phys. C} \textbf{71}, 415 (1996).
\bibitem{benchmark1}
    M. Battaglia \textit{et al.}, Post-LEP CMSSM benchmarks for supersymmetry,
    \texttt{hep-ph/0112013}.
\bibitem{benchmark2}
    N. Ghodbane and H.-U. Martyn, Compilation of SUSY particle spectra from
    Snowmass 2001 benchmark models, \texttt{hep-ph/0201233}.
\bibitem{benchmark3}
    M. Battaglia, A. De Roeck, J. Ellis, F. Gianotti, K.A. Olive and L. Pape,
    \textit{Eur. Phys. J. C} \textbf{33}, 273 (2004).
\bibitem{benchmark4}
    J. Ellis, S. Heinemeyer, K.A. Olive, G. Weiglein, \textit{JHEP} \textbf{0605}, 005 (2006).
\bibitem{us} A.V. Gladyshev, D.I. Kazakov and M.G. Paucar,
    \textit{Mod. Phys. Lett. A} \textbf{20}, 3085 (2005).
\bibitem{Anom1}
    G.W. Bennett \textit{et al.},
    \textit{Phys. Rev. Lett.} \textbf{92}, 161802 (2004).
\bibitem{Anom2}
    M. Davier, S. Eidelman, A. Hocker and Z. Zhang,
    \textit{Eur. Phys. J. C} \textbf{31}, 503 (2003).
\bibitem{Anom3}
    K. Hagiwara, A.D. Martin, D. Nomura and T.Teubner,
    \textit{Phys. Rev. D} \textbf{69}, 093003 (2004).
\bibitem{Anom4}
    J.F. de Troconiz and F.J. Yndurain,
    \textit{Phys. Rev. D} \textbf{71}, 073008 (2005).
\bibitem{Anom5}
    K. Melnikov and A. Vainshtein,
    \textit{Phys. Rev. D} \textbf{70}, 113006 (2004).
\bibitem{Anom6}
    M. Passera,
    \textit{J. Phys. G} \textbf{31}, R75 (2005).
\bibitem{Anom7}
    W. de Boer, M.Huber, C.Sander and D.I. Kazakov,
    \textit{Phys. Lett. B} \textbf{515}, 283 (2001).
\bibitem{Anom8}
    W. de Boer, M. Huber, C. Sander, A.V. Gladyshev and D.I. Kazakov,
    A global fit to the anomalous magnetic moment, $b \to X_s \gamma$
    and Higgs limits in the constrained MSSM, in \textit{Supersymmetry
    and unification of fundamental interactions: Proceedings},
    eds. D.I.~Kazakov and A.V. Gladyshev (World Scientific, 2002), p. 196,
    \texttt{hep-ph/0109131}.
\bibitem{Unif1}
    J.R. Ellis, S. Kelley and D.V. Nanopoulos,
    \textit{Phys. Lett. B} \textbf{260}, 131 (1991).
\bibitem{Unif2}
    U. Amaldi, W. de Boer and H. Furstenau,
    \textit{Phys. Lett. B} \textbf{260}, 447 (1991).
\bibitem{Unif3}
    C. Giunti, C.W. Kim and U.W. Lee,
    \textit{Mod. Phys. Lett. A} \textbf{6}, 1745 (1991).
\bibitem{LSP1}
    J.R. Ellis \textit{et al.},
    \textit{Nucl. Phys. B} \textbf{238}, 453 (1984).
\bibitem{LSP2}
    G. Jungman, M. Kamionkowski and K. Griest,
    \textit{Phys. Rep.} \textbf{267}, 195 (1996).
\bibitem{higgslimits1}
    LEP Higgs Working Group for Higgs boson searches, OPAL
    Collaboration, ALEPH Collaboration, DELPHI Collaboration and
    L3 Collaboration, Search for the Standard Model Higgs Boson at
    LEP, CERN-EP/2003-011, \texttt{hep-ex/0306033}.
\bibitem{higgslimits2}
    Joint LEP2 SUSY Working Group, Combined LEP Chargino
    Results up to 208 GeV,\\
    \verb+http://lepsusy.web.cern.ch/lepsusy/www/inos_moriond01/charginos_pub.html+
\bibitem{higgslimits3}
    Combined LEP Selectron/Smuon/Stau Results, 183-208 GeV,\\
    \verb+http://lepsusy.web.cern.ch/lepsusy/www/sleptons_summer02/slep_2002.html+
\bibitem{ISAJET}
    H. Baer, F. Paige, S.D. Protopescu and X. Tata,
    ISAJET 7.69. A Monte Carlo Event Generator for
    $\it{pp, p\bar{p}}$ and $\it{e^{+},e^{-}}$ reactions, \texttt{hep-ph/0312045}.
\bibitem{kraml1}
    G. Belanger, S. Kraml, A. Pukhov, \texttt{hep-ph/0502079}.
\bibitem{kraml2}
    B.C. Allanach, S. Kraml, W. Porod, \textit{JHEP} \textbf{03}, 016(2003).
\bibitem{kraml3}
    \verb"http://cern.ch/kraml/comparison/"
\bibitem{WMAP1}
    C.L. Bennett \textit{et al.},
    \textit{Astrophys. J. Suppl.} \textbf{148}, 1 (2003).
\bibitem{WMAP2}
    D.N. Spergel \textit{et al.},
    \textit{Astrophys. J. Suppl.} \textbf{148}, 175 (2003).
\bibitem{micromegas1}
    G. Belanger, F. Boudjema, A. Pukhov and A. Semenov,
    MicrOMEGAs: Version 1.3, \texttt{hep-ph/0405253}.
\bibitem{micromegas2}
    G. Belanger, F. Boudjema, A. Pukhov and A. Semenov,
    \textit{Comput. Phys. Commun.} {\bf 149}, 103 (2002).
\bibitem{stopprod1}
    B.C. Allanach \textit{et al}., Les Houches physics at TeV colliders 2005
    beyond the standard model working group: Summary report, \texttt{hep-ph/0602198}.
\bibitem{stopprod2}
    A.M. Abazov \textit{et al}., \textit{Phys. Lett. B} \textbf{645}, 119 (2007).
     [hep-ex/0611003]
\bibitem{stopdecay}
    K. Hikasa and M. Kobayashi, \textit{Phys. Rev. D} \textbf{36}, 724 (1987).
\end{thebibliography}
\end{document}